\documentclass{PoS}
\usepackage{epsfig,psfig}

\title{Critical phenomena and renormalization-group
flow of multi-parameter $\Phi^4$ field theories}

\ShortTitle{RG flow of multi-parameter $\Phi^4$ field theories}

\author{\speaker{Ettore Vicari}\thanks{ 
I acknoledge the important
contributions of my collaborators, in particular Andrea Pelissetto, on the
topics discussed in this talk.  }
  \\ Dipartimento di Fisica, Universit\'a di Pisa and INFN\\
  E-mail: \email{vicari@df.unipi.it}}
      
\abstract{ 
  
  In the framework of the renormalization-group (RG) approach, critical
  phenomena can be investigated by studying the RG flow of multi-parameter
  $\Phi^4$ field theories with an $N$-component fundamental field, containing
  up to 4th-order polynomials of the field.  Some physically interesting
  systems require $\Phi^4$ field theories with several quadratic and quartic
  parameters, depending essentially on their symmetry and symmetry-breaking
  pattern at the transition.  Results for their RG flow apply to disorder
  and/or frustrated systems, anisotropic magnetic systems, density wave
  models, competing orderings giving rise to multicritical behaviors.
  
  The general properties of the RG flow in multi-parameter $\Phi^4$ field
  theories are discussed. An overview of field-theoretical results for some
  physically interesting cases is presented, and compared with other
  theoretical approaches and experiments.  Finally, this RG approach is
  applied to investigate the nature of the finite-temperature transition of
  QCD with $N_f$ light quarks.

}

\FullConference{The XXV International Symposium on Lattice Field Theory\\
                 July 30-4 August 2007\\
                 Regensburg, Germany}

\begin{document}

\section{Introduction}
\label{lsec-intro}

In the framework of the renormalization-group (RG) approach to critical
phenomena, a quantitative description of many continuous phase transitions can
be obtained by considering an effective Landau-Ginzburg-Wilson (LGW) $\Phi^4$
field theory, containing up to fourth-order powers of the field components.
The simplest example is the O($N$)-symmetric $\Phi^4$ theory, defined by the
Lagrangian density
\begin{equation}
{\cal L}_{O(N)} =  
{1\over 2} \sum_i (\partial_\mu \Phi_{i})^2 + 
{1\over 2} r \sum_i \Phi_{i}^2  + 
{1\over 4!} u (\sum_{i} \Phi_i^2 )^2
\label{HON}
\end{equation}
where $\Phi$ is an $N$-component real field.  These $\Phi^4$ theories describe
phase transitions characterized by the symmetry breaking
O($N$)$\rightarrow$O($N-1$).  We mention the Ising universality class for
$N=1$ (which is relevant for the liquid-vapor transition in simple fluids, for
the Curie transition in uniaxial magnetic systems, etc...), the $XY$
universality class for $N=2$ (which describes the superfluid transition in
$^4$He, transitions in magnets with easy-plane anisotropy, and in
superconductors), the Heisenberg universality class for $N=3$ (it describes
the Curie transition in isotropic magnets).  Moreover, the limit $N\rightarrow
0$ describes the behavior of dilute homopolymers in a good solvent in the
limit of large polymerization.  See, e.g., Refs.~\cite{PV-r,ZJ-book} for
recent reviews.

Beside the transitions described by O($N$) models, there are also other
physically interesting transitions described by more general
Landau-Ginzburg-Wilson (LGW) $\Phi^4$ field theories, characterized by more
complex symmetries and symmetry breaking patterns.  The general LGW $\Phi^4$
theory for an $N$-component field $\Phi_i$ can be written as
\begin{equation}
{\cal L} = 
{1\over 2} \sum_i (\partial_\mu \Phi_{i})^2 + 
{1\over 2} \sum_i r_i \Phi_{i}^2  + 
{1\over 4!} \sum_{ijkl} u_{ijkl} \; \Phi_i\Phi_j\Phi_k\Phi_l  
\label{generalH}
\end{equation}
where the number of independent parameters $r_i$ and $u_{ijkl}$ depends on the
symmetry group of the theory.  Here, we are only assuming a parity symmetry
which forbids third-order terms.  In the field-theoretical (FT) approach the
RG flow is determined by a set of RG equations for the correlation functions
of the order parameter.  This approach has been applied to investigate the
critical behavior of disorder and/or frustrated systems, magnets with
anisotropy, spin and density wave models, competing orderings giving rise to
multicritical behaviors, and also the finite-temperature transition in
hadronic matter.

The main issue discussed in this talk is the RG flow of general
multi-parameter $\Phi^4$ field theories, and its applications to the study of
critical phenomena in statistical systems. In particular, we consider FT
perturbative approaches based on expansions in powers of renormalized quartic
couplings.  An overview of FT results for physically interesting cases is
presented and compared with experiments and other approaches such as lattice
techniques.  Finally, this RG approach is applied to the study of the nature
of the finite-temperature transition in QCD with $N_f$ light flavours.

\section{Renormalization-group theory of critical phenomena}
\label{RGth}

\begin{figure}
\begin{center}
{\begin{tabular}{cc}
\psfig{width=5.5truecm,angle=0,file=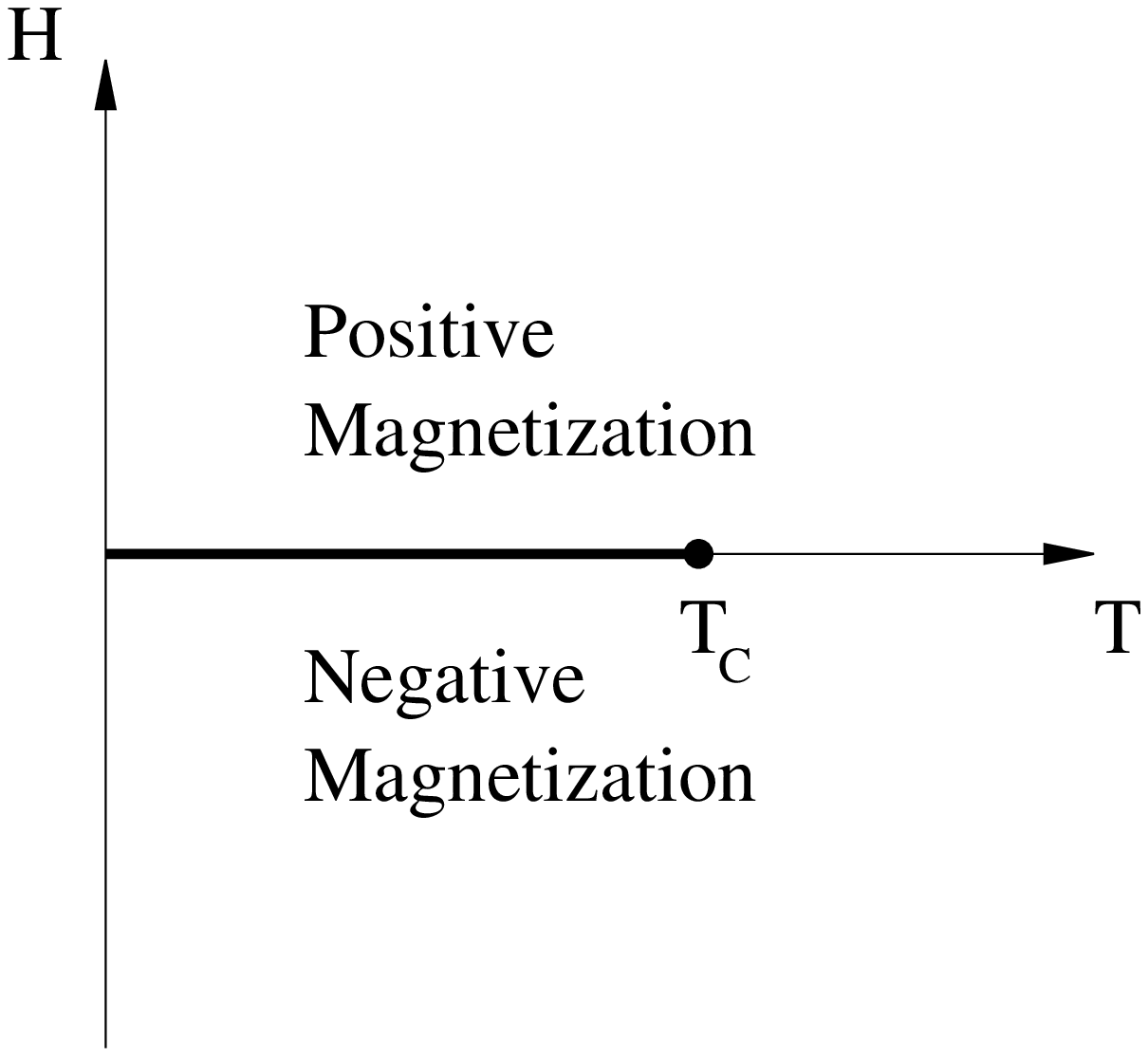} &
\hskip 0.1truecm
\psfig{width=5.5truecm,angle=0,file=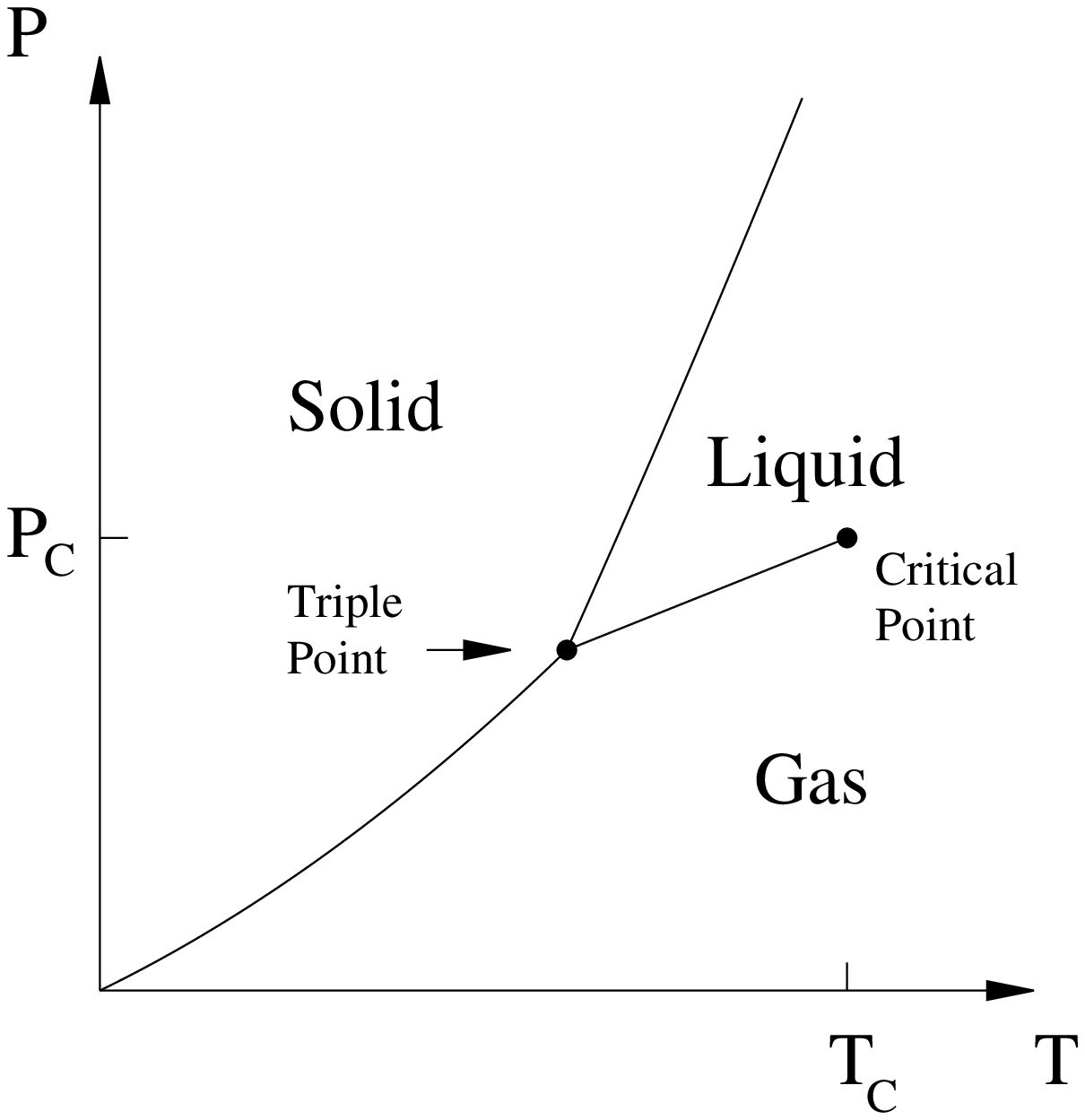} 
\end{tabular}}
\caption{Typical phase diagrams of ferromagnetic materials (left) and liquids (right)}
\label{pdml}
\end{center}
\end{figure}

Critical phenomena are observed in many physical systems when they undergo
second-order (continuous) transitions characterized by a nonanalytic behavior
due to a diverging length scale.~\footnote{ Continuous transitions are
  generally characterized power-law behaviors.  For example, defining the
  reduced temperature as $t\equiv T/T_c-1$, in the disordered phase the
  magnetic susceptibility $\chi$ and the correlation length $\xi$ diverge as
  $\chi\sim t^{-\gamma}$ and $\xi\sim t^{-\nu}$ respectively, the specific
  heat behaves as $C_H\sim |t|^{-\alpha}$, the two-point function $W^{(2)}(x)$
  decays as $1/x^{d-2+\eta}$ at $T=T_c$, the magnetization vanishes as $M\sim
  (-t)^{\beta}$ in the ordered phase, etc...  In standard continuous
  transitions only two critical exponents are independent, because there are
  several scaling relations: $\gamma=(2-\eta)\nu$, $\alpha=2-d\nu$, $\beta=\nu
  (d-2+\eta)/2$.}  Classical examples are transitions in ferromagnetic
materials and liquids, whose phase diagrams are sketched in Fig.~\ref{pdml}.
The first general framework to understand critical phenomena was proposed by
Landau~\cite{Landau-37}; it was essentially a mean-field approximation.  A
satisfactory understanding was later achieved by the Wilson RG
theory~\cite{Wilson-71}.

The main ideas to describe the critical behavior at a continuous transition
are (i) the existence of an order parameter which effectively describes the
critical modes; (ii) the scaling hypothesis: singularities arise from the
long-range correlations of the order parameter, which develop a diverging
length scale; (iii) universality: the critical behavior is essentially
determined by a few global properties, such as the space dimensionality, the
nature and the symmetry of the order parameter, the symmetry breaking, the
range of the effective interactions. The RG theory of critical
phenomena~\cite{Wilson-71,WK-72,Fisher-74} provides a general framework where
these features naturally arise. It considers a RG flow in a Hamiltonian space.
The critical behavior is associated with a fixed point (FP) of the RG flow
where only a few perturbations are relevant.  The corresponding positive
eigenvalues of the linearized theory around the FP are related to the critical
exponents $\nu$, $\eta$, etc...

According to the RG theory, the singular part of the Gibbs free energy obeys a
scaling law
\begin{eqnarray}
{\cal F}_{\rm sing}(u_1,u_2,\ldots,u_k,\ldots) = 
b^{-d} {\cal F}_{\rm sing}(b^{y_1} u_1,b^{y_2} u_2,\ldots,
   b^{y_k} u_k,\ldots)
\label{freeen}
\end{eqnarray}
where $u_i$ are the nonlinear scaling fields, which are analytic functions of
the model parameters.  In a standard continuous transition there are two
relevant scaling fields with $y_i>0$, $u_t$ and $u_h$, and an infinite set of
irrelevant fields $w_i$ with $y_i < 0$.  The relevant fields may be identified
with the reduced temperature $t\equiv T/T_c-1$ and the external (magnetic)
field $H$, i.e.  $u_t\sim t$ and $u_h\sim H$ for $t,H\to 0$.  Setting $b^{y_t}
|u_t| = 1$, we can write
\begin{equation}
{\cal F}_{\rm sing} = |u_t|^{d/y_t} 
{\cal F}_{\rm sing}(u_h |u_t|^{-y_h/ y_t},w_i |u_t|^{-y_i/ y_t}) 
\label{freen2}
\end{equation}
Since $w_i |u_t|^{-y_i/ y_1}\to0$ for $t\to 0$, we can expand with respect to
the arguments containing the irrelevant scaling fields, obtaining
\begin{eqnarray} 
 {\cal F}_{\rm sing}\approx
  |t|^{d/y_t} f(|H| |t|^{-y_h/ y_t}) 
  + |t|^{d/y_t+\Delta_i} f_{(1,i)}(|H| |t|^{-y_h/ y_t}) + ... 
\label{freeen3}
\end{eqnarray}
where $y_t = {1/ \nu}$, $y_h = (d+2-\eta)/2$, $\Delta_i=-y_i/y_t>0$.  The
first term of the r.h.s. of Eq.~(\ref{freeen3}) is the universal asymptotic
critical behavior, while the other terms give rise to nonuniversal scaling
corrections.  The above scaling relations can be easily extended to allow for
finite-size systems.  In the presence of other relevant perturbations beside
$t$ and $H$, one observes more complicated multicritical behaviors.

\section{Field-theoretical perturbative approach}
\label{FTapproach}

The RG theory provides the basis for the FT approaches to the study of
critical phenomena. The critical behavior can be determined by the RG flow of
a corresponding Euclidean quantum field theory (QFT).

\subsection{Critical behavior of statistical systems and RG flow of 
Euclidean QFTs}
\label{QFT}

Let us consider the Ising model defined on a $d$-dimensional lattice:
\begin{equation}
H = - J \sum_{\langle xy\rangle} \sigma_x \sigma_y, \quad
\sigma_x=\pm 1, \quad Z = \sum_{\{\sigma_x\}} {\rm exp} (-H/T),
\label{ising}
\end{equation}
where the sum in the Hamiltonian is over the nearest-neighbor sites of the
lattice.  The critical behavior is due to the long-range modes, with $l\gg 1$.
In order to describe these critical modes, one may perform blocking averages
over sizes $a\ll l$, and derive an effective Hamiltonian for the block average
variables, which can be considered as real variables $\varphi_x$ attached to
the blocks. The simplest effective Hamiltonian of the block variables
$\varphi_x$ which preserves the ${\mathbb Z}_2$ symmetry may be written as
\begin{equation}
H_{\rm eff} = a^{d-2}\sum_{x,\mu}
(\varphi_{x+\mu} - \varphi_x)^2 + u a^d \sum_x ( \varphi_x^2 - v^2)^2
\label{latphi4}
\end{equation}
This blocking procedure does not affect the long range modes at the scale $l$,
and therefore it does not change the universality class of the transition.
Then, we may consider the limit $a\rightarrow 0$ of $H_{\rm eff}$:
\begin{equation}
H_{\varphi^4} =  \int d^d x {\cal H}(\varphi), \qquad
{\cal H}(\varphi)=
{1\over 2}(\partial_\mu \varphi)^2 + {1\over 2} r\varphi^2 
+ {1\over 4!} u \varphi^4 
\label{phi4cont}
\end{equation}
where $r-r_c\propto T - T_c$. Again, this limit does not change the
universality class. The corresponding partition function
\begin{equation}
Z = \int [d\varphi] \exp[ - \int d^d x {\cal H}(\varphi)]
\label{contpartf}
\end{equation}
is the path integral of an Euclidean QFT with ${\cal L}(\varphi)={\cal
  H}(\varphi)$.  Therefore, the critical behavior of the original Ising model
is related to the behavior of the correlation functions of the $\Phi^4$ QFT
for a real one-component field in the massless limit.  Analogous euristic
arguments can be applied to other statistical systems and corresponding QFTs.

Note that the way back provides a nonperturbative formulation of an Euclidean
QFT, from the critical behavior of a statistical model.  An important example
is the Wilson lattice formulation of QCD~\cite{Wilson-74} defined in the
critical (continuum) limit of a 4D statistical model.

\subsection{Perturbative schemes in the field-theory approach}
\label{perft}

A successful approach to the study of critical phenomena 
exploits the relation with the RG flow of a corresponding QFT, and
therefore the typical techniques used in QFTs.  We are interested in the
critical behavior of the ``bare'' correlation functions $\Gamma_n(p; r, u,
\Lambda)$, which correspond to the ``physical'' correlation functions of the
statistical system, unlike high-energy physics where the physical correlation
functions are the renormalized ones.

In this section we consider the simplest case of the O($N$) $\Phi^4$ field
theory (\ref{HON}).  For $d<4$, i.e. $d=3,2$, the theory is
super-renormalizable since the number of primitively divergent diagrams is
finite.  Such divergences are related to the necessity of performing an
infinite renormalization of the parameter $r$ appearing in the bare
Lagrangian, see, e.g., the discussion in Ref.~\cite{BB-85}. This problem can
be avoided by replacing the quadratic parameter $r$ of the Lagrangian with the
mass $m$ (inverse second-moment correlation length) defined by
\begin{equation}
m^{-2} = \, {1\over \Gamma^{(2)}(0)} \, 
     \left. {\partial \Gamma^{(2)}(p^2) \over \partial p^2}\right|_{p^2=0},
\end{equation}
where $ \Gamma^{(2)}_{ab}(p^2) =\Gamma^{(2)}(p^2)\delta_{ab}$ is the
one-particle irreducible two-point function.  Perturbation theory in terms of
$m$ and the bare quartic coupling $u$ is finite in $d<4$.  In the massive
zero-momentum (MZM) scheme,~\cite{Parisi-80} one considers a set of
zero-momentum renormalization conditions in the massive (disordered) phase,
\begin{eqnarray}
\Gamma^{(2)}(p) = Z_\phi^{-1} \left[ m^2+p^2+O(p^4)\right],
\qquad \Gamma^{(4)}(0) = Z_\phi^{-2} m^{4-d} g
\label{rgcond}
\end{eqnarray}
where $\Gamma^{(n)}$ are one-particle irreducible correlation functions, and
$g$ is the renormalized MZM coupling.  Moreover, one defines
$Z_t^{-1}=\Gamma^{(1,2)}(0)$, where $\Gamma^{(1,2)}(p)$ is the one-particle
irreducible two-point function with an insertion of ${1\over2}\phi^2$.  The
critical limit $m\rightarrow 0$ can be studied by writing Callan-Symanzik RG
equations for the renormalized correlation functions $\Gamma^{(n)}_r =
Z_\varphi^{n/2} \Gamma^{(n)} $,
\begin{eqnarray}
\left[ m {\partial\over \partial m} +
\beta(g) {\partial\over \partial g} - \frac{1}{2} n \eta_\varphi(g)\right]
\Gamma^{(n)}_r(p)
= [2-\eta_\varphi(g)] m^2 \Gamma^{(1,n)}_r(p;0),
\label{CSeq}
\end{eqnarray}
see, for example, Ref.~\cite{ZJ-book}.  The RG functions
\begin{equation}
\beta(g) = m {\partial g\over \partial m},\qquad
\eta_{\phi,t}(g) = {\partial {\rm ln} Z_{\phi,t}\over \partial {\rm
ln} m}
\label{rgfuncdef}
\end{equation}
can be computed as power series of $g$. For the 3D O($N$) $\Phi^4$ models they
have been computed to six and seven loops respectively~\cite{BNGM-77}.  When
$m\rightarrow 0$ the coupling $g$ is driven toward an infrared-stable fixed
point (FP), i.e. a zero $g^*$ of the $\beta$-function, $\beta(g) = - \omega
(g^*-g) + O[(g^*-g)^x]$ with $x>1$.  The Callan-Symanzik
$\beta$-function $\beta(g)$ is nonanalytic at
$g=g^*$~\cite{PV-98}.  Using the RG equations, one can also identify $\eta =
\eta_\phi(g^*)$ and $1/\nu=2-\eta_\phi(g^*)+\eta_t(g^*)$.

One may also consider an alternative perturbative scheme: the $\overline{\rm
  MS}$ renormalization scheme~\cite{tHV-72}, defined at $T=T_c$, i.e. in the
massless theory.  This is based on the dimensional regularization and the
subtraction of the $1/\epsilon$ poles ($\epsilon\equiv 4-d$) to obtain the
renormalized correlation functions.  One sets $\Phi = [Z_\phi(g)]^{1/2}
\Phi_r$, $u = A_d \mu^\epsilon Z_g(g)$, where $A_d$ is a $d$-dependent
constant, and $g$ is the $\overline{\rm MS}$ renormalized quartic coupling.
Moreover, one defines a mass renormalization constant $Z_t$ by requiring $Z_t
\Gamma^{(1,2)}$ to be finite when expressed in terms of $g$.  Then one derives
the RG functions $\beta(g) = \mu {\partial g/\partial \mu}$
and  $\eta_{\phi,t}(g)
= {\partial \ln Z_{\phi,t}/ \partial \ln \mu}$.  Analogously to the MZM
scheme, the nontrivial zero $g^*$ of the $\beta$-function is the stable FP,
and the critical exponents are obtained by evaluating the RG functions
$\eta_{\phi,t}(g)$ at $g=g^*$.  Within this scheme, one can perform the
so-called $\epsilon\equiv 4-d$ expansion~\cite{WF-72}, i.e. and expansion
about $d=4$. Alternatively, one can fix $d$ after
renormalization~\cite{Dohm-85}, for example $d=3$, obtaining a fixed-dimension
expansion in the $\overline{\rm MS}$ coupling $g$.

FT perturbative expansions are divergent.  If we consider a quantity
$S(g)$ that has a perturbative expansion $S(g)
\approx \sum s_k g^k$, the large-order behavior of the coefficients is
given by $s_k \sim k!
\,(-a)^{k}\, k^b \,\left[ 1 + O(k^{-1})\right]$, with $a>0$ for $d<4$.
Thus, in order to obtain accurate results, an appropriate resummation
is required before evaluating the RG functions at the fixed-point
value $g^*$.  This can be done by exploiting their Borel summability,
which has been proved for the fixed-dimension expansion 
in $d<4$, see e.g. Refs.~\cite{ZJ-book,PV-r} and
references therein, and has been conjectured for the $\epsilon$
expansion. Moreover, one can also exploit knowledge of the high-order
behavior of the expansion, which is computed by semiclassical
instanton calculations, see, e.g., Refs.~\cite{LZ-77,ZJ-book}.  
Note that the results of this resummation are essentially nonpertubative,
because it uses nonperturbative information.

\subsection{The 3D Ising and $XY$ universality classes}
\label{onmodels}

\begin{table}
\begin{center}
\begin{tabular}{llllll}
\hline
\multicolumn{2}{c}{3D Ising exponents}& 
\multicolumn{1}{c}{$\nu$}& 
\multicolumn{1}{c}{$\alpha$}& 
\multicolumn{1}{c}{$\eta$}& 
\multicolumn{1}{c}{$\beta$}\\
\hline  
EXPT & liquid-vapour  & 0.6297(4) & 0.111(1) & 0.042(6) & 0.324(2) \\
& fluid mixtures & 0.6297(7) & 0.111(2) & 0.038(3) & 0.327(3) \\
     & uniaxial magnets & 0.6300(17) & 0.110(5) & & 0.325(2) \\

PFT  & 6,7-$l$ MZM \cite{GZ-98} & 0.6304(13) & 0.109(4) &  
0.034(3) & 0.326(1) \\

& $O(\epsilon^5)$ exp \cite{GZ-98}& 0.6290(25) & 0.113(7) & 
0.036(5) & 0.326(3) \\ 

Lattice  & HT exp \cite{CPRV-02} & 0.63012(16) & 0.1096(5) & 
0.0364(2) & 0.3265(1) \\
         & MC \cite{DB-03} & 0.63020(12) & 0.1094(4) & 0.0368(2) & 0.3267(1) \\
\hline
\end{tabular}
\caption{
Estimates of the critical exponents of the 3D Ising universality
class, from experiments (taken from the review \cite{PV-r}),
resummation of the FT 6,7-loop calculations within the MZM scheme and
of $O(\epsilon^5)$ expansions, and from lattice techniques: 25th order
high-temperature (HT) expansion and Monte Carlo (MC) simulations.  }
\label{isingres}
\end{center}
\end{table}

The effectiveness of the RG approach based on $\Phi^4$ QFTs can be
appreciated by looking at the results for the 3D O($N$)
models, and in particular for the Ising and $XY$
universality classes.

The Ising universality class corresponds to a $\Phi^4$ theory with a real
one-component field. It describes transitions in several physical systems,
such as liquid-vapor systems, fluid mixtures, uniaxial magnets.
Table~\ref{isingres} shows selected results from experiments,
field-theoretical and lattice computations.  A more complete review of results
can be found in Ref.~\cite{PV-r}.  FT results are quite precise and in good
agreement with experiments and lattice computations, which provide the most
precise theoretical estimates.

The 3D $XY$ universality class is characterized by a two-component order
parameter and the symmetry $U(1)$. It corresponds to the O($N$) model
(\ref{HON}) with $N=2$.  An interesting representative of this universality
class is the superfluid transition of $^4$He along the $\lambda$-line
$T_\lambda(P)$, where the quantum amplitude of helium atoms is the order
parameter.  The superfluid transition of $^4$He provides an exceptional
opportunity for a very accurate experimental test of the RG predictions.
Moreover, experiments in a microgravity environment, for instance on the Space
Shuttle \cite{Lipa-etal-96}, can achieve a significant reduction of the
gravity-induced broadening of the transition.  Exploiting these favorable
conditions, the specific heat of liquid helium was measured to within a few nK
from the $\lambda$-transition~\cite{Lipa-etal-96}.  In Table~\ref{XYres} we
show selected results from experiments, field-theoretical and lattice
computations.  We note that there is a significant difference between the
experimental result and the best theoretical estimates using lattice
techniques.  FT results are in good agreement with experiments and lattice
computations, but they are not sufficiently precise to distinguish them.  This
discrepancy should not be necessarily considered as a failure of the RG theory,
however it calls for further investigations.  A proposal of a new space
experiment has been presented in Ref.~\cite{LWNA-05}.

\begin{table}
\begin{center}
\begin{tabular}{lllll}
\hline
\multicolumn{2}{c}{3D $XY$ exponents}& 
\multicolumn{1}{c}{$\alpha$}& 
\multicolumn{1}{c}{$\nu$}& 
\multicolumn{1}{c}{$\eta$}\\
\hline  
EXPT & $^4$He \cite{Lipa-etal-96} & $-$0.0127(3) & 0.6709(1)$^*$ &  \\
PFT  & 6,7-$l$ MZM \cite{GZ-98} & $-$0.011(4) & 0.6703(15) & 0.035(3) \\
     & $O(\epsilon^5)$ exp \cite{GZ-98} & $-$0.004(11) & 0.6680(35) & 0.038(5) \\
Lattice  & MC+HT \cite{CHPV-06} & $-$0.0151(3)$^*$ & 0.6717(1) & 0.0381(2) \\
          &  MC \cite{BNPS-06}  & $-$0.0151(9)$^*$ & 0.6717(3) & \\
\hline
\end{tabular}
\caption{
Results for the critical exponents of the 3D $XY$ universality class,
from experiments on $^4$He,  
resummation of the FT 6,7-loop calculations within the MZM scheme and
of $O(\epsilon^5)$ expansion, and from lattice techniques:
a sinergy of HT expansions and MC simulations (MC+HT)
and MC simulations (MC).
Results marked by an asterisk are obtained by using the
hyperscaling relation $\alpha=2-d\nu$.
}
\label{XYres}
\end{center}
\end{table}

\section{The RG flow in  multi-parameter 
  $\Phi^4$ field theories and the $\eta$ conjecture}
\label{rgfloweta}

Beside transitions belonging to the O($N$) universality classes, many other
physically interesting transitions are described by more general
multi-parameter $\Phi^4$ field theories, cf. Eq.~(\ref{generalH}),
characterized by more complex symmetries~\cite{Aharony-76,BLZ-76,PV-r}.  The
number of independent parameters $r_i$ and $u_{ijkl}$ depends on the symmetry
group of the theory. An interesting class of models are those in which $\sum_i
\Phi^2_i$ is the unique quadratic polynomial invariant under the symmetry
group of the theory, corresponding to the case all field components become
critical simultaneously.  This requires that all $r_i$ are equal, $r_i = r$,
and $u_{ijkl}$ must be such not to generate other quadratic invariant terms
under RG transformations, for example, it must satisfy the trace
condition~\cite{BLZ-74} $\sum_i u_{iikl} \propto \delta_{kl}$.  In these
models, criticality is driven by tuning the single parameter $r$, which
physically may correspond to the reduced temperature.  More general LGW
$\Phi^4$ theories, which allow for the presence of independent quadratic
parameters $r_i$, must be considered to describe multicritical behaviors where
there are independent correlation lengths that diverge simultaneously, which
may arise from the competition of distinct types of ordering.
Note that, like the simplest O($N$) models (\ref{HON}), all
multi-parameter $\Phi^4$ field theories are expected to be trivial in four
dimensions.

\begin{figure}[!bt]
\begin{tabular}{cc}
\psfig{width=5.5truecm,angle=0,file=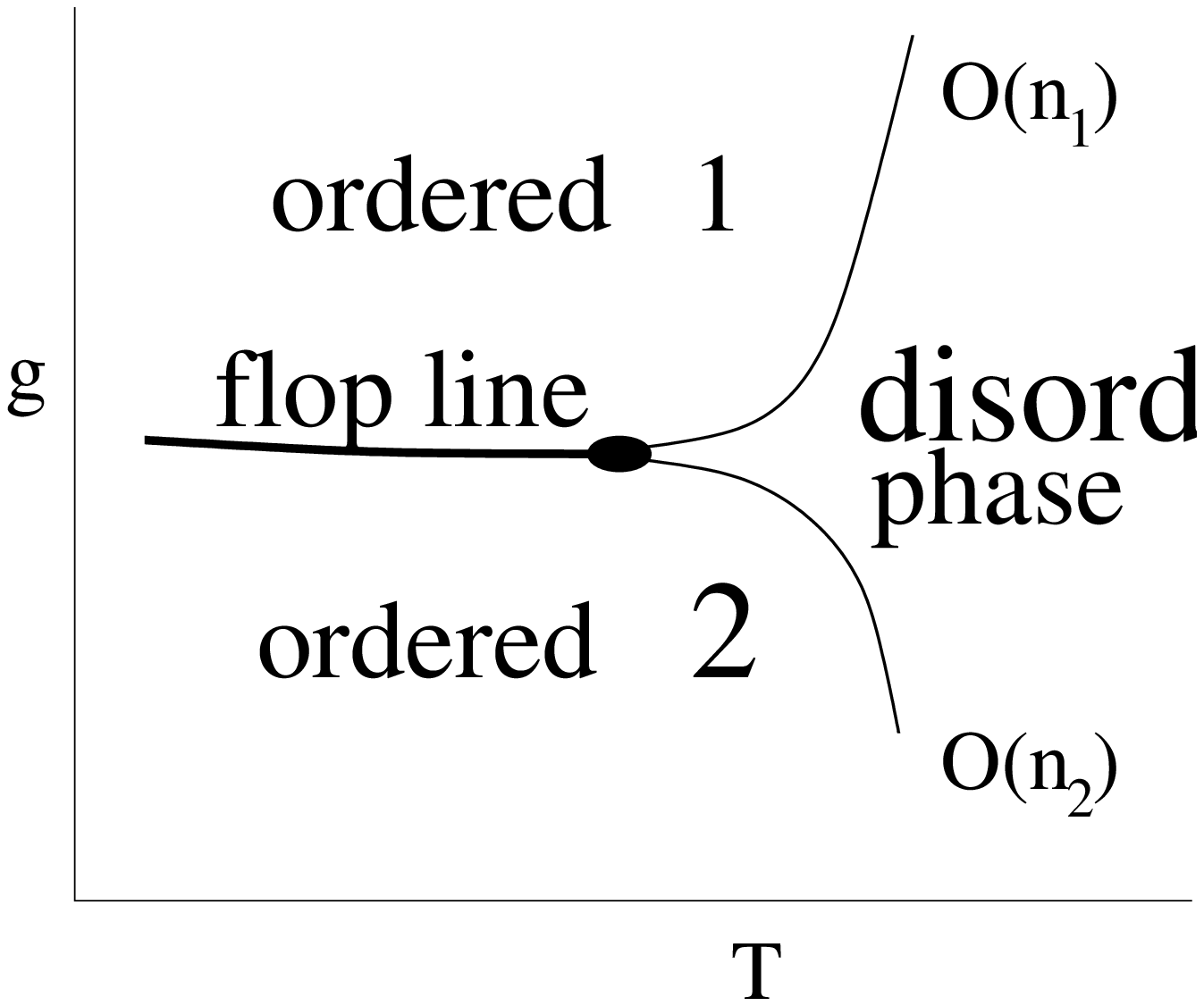} &
\hskip 2truecm
\psfig{width=5.5truecm,angle=0,file=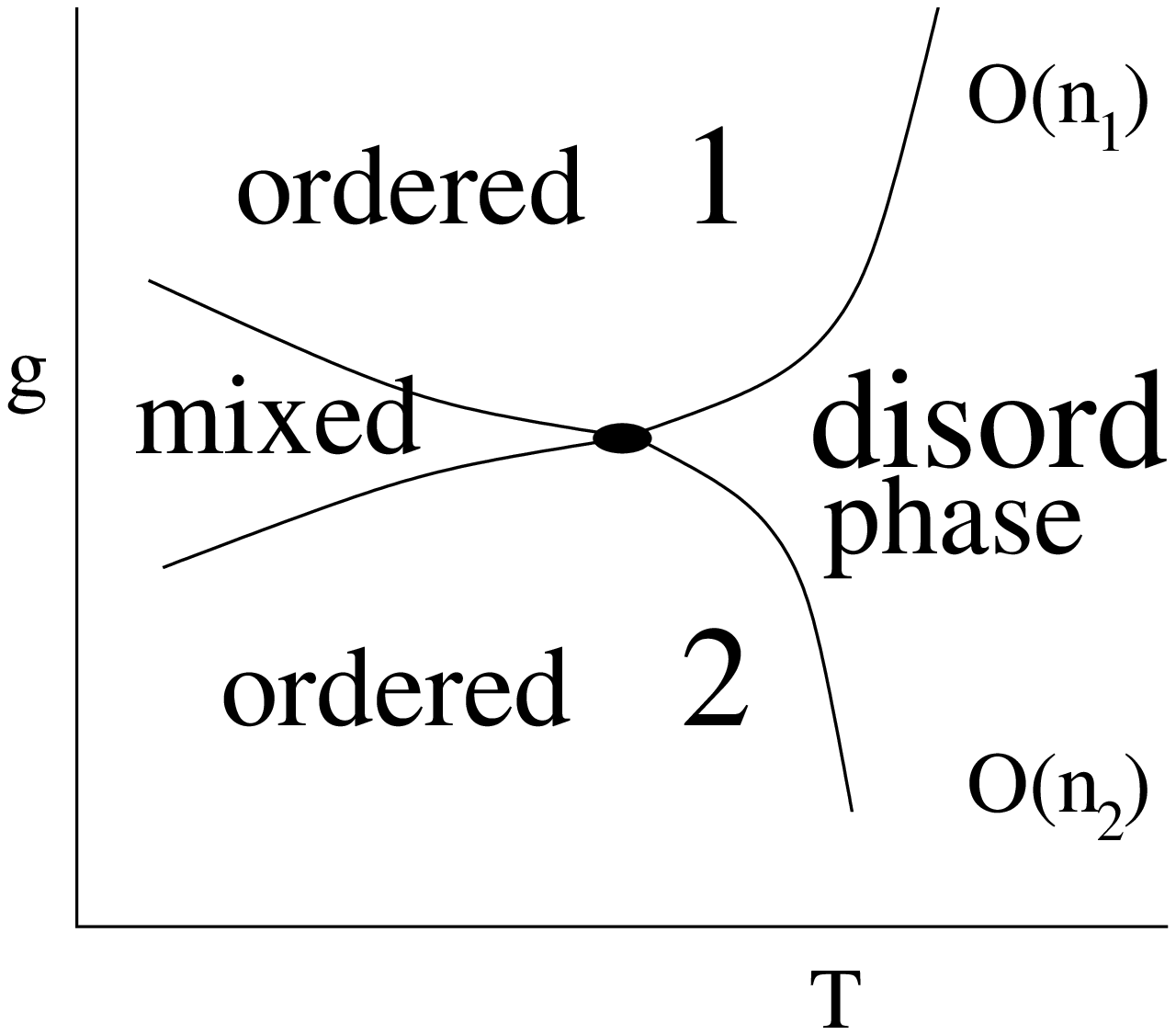} 
\end{tabular}
\vspace{2mm}
\caption{
  Phase diagrams with multicritical points where different transition lines
  meet.  }
\label{mcps}
\end{figure}

\subsection{Physically interesting multi-parameter $\Phi^4$ field theories}
\label{physint}

Some physically interesting examples of
multi-parameter $\Phi^4$ field theories are:

\begin{itemize}

\item[$\bullet$]
The O($M$)$\otimes$O($N$) models, which is the most general $\Phi^4$ theory
with a $N\times M$ matrix fundamental field $\Phi_{ai}$ 
($a=1,...,M$, $i=1,...,N$), and the symmetry O($M$)$\otimes$O($N$), 
\begin{eqnarray}
{\cal L} = 
\sum_{ai} [ (\partial_\mu \Phi_{ai})^2 + r \Phi_{ai}^2 ]
+ u_0 ( \sum_{ai} \Phi_{ai}^2 )^2
 + v_0 [\sum_{a,b} ( \sum_i \Phi_{ai}\Phi_{bi} )^2 
-( \sum_{ai} \Phi_{ai}^2 )^2]
\label{omnmo}
\end{eqnarray}
They are relevant for transitions in noncollinear frustrated
magnets and in $^3$He. See Sec.~\ref{omnphi4th}.

\item[$\bullet$]
The $MN$ models with a real $M\times N$ matrix field $\Phi_{ai}$: 
\begin{eqnarray}
{\cal L}=\sum_{i,a}\left[ (\partial_\mu \Phi_{ai})^2 + 
         r \Phi_{ai}^2 \right] 
+\sum_{ij,ab} \left( u_0 + v_0 \delta_{ij} \right)
          \Phi^2_{ai} \Phi^2_{bj} 
\label{mnmo}
\end{eqnarray}
They describe transitions in randomly diluted $M$-component spins for
$N\rightarrow 0$, see below in Sec.~\ref{rdis}, and magnets with cubic
anisotropy for $M=1$, $N=2,3$, see e.g. Refs.~\cite{Aharony-76,CPV-00}.

\item[$\bullet$]
The spin-density wave model has two complex $N$-component 
order-parameter fields:
\begin{eqnarray}
{\cal L}=&&\vert\partial_\mu \Phi_{1}\vert^2 +
                 \vert\partial_\mu \Phi_{2}\vert^2 
+r(\vert\Phi_{1}\vert^2+\vert\Phi_{2}\vert^2 )
+u_{1,0} (\vert\Phi_{1}\vert^4+\vert\Phi_{2}\vert^4)
\nonumber\\&&
+u_{2,0}(\vert\Phi_{1}^2\vert^2+\vert\Phi_{2}^2\vert^2)
   +w_{1,0}\vert\Phi_{1}\vert^2 \vert\Phi_{2}\vert^2
   +w_{2,0}\vert\Phi_{1}\cdot \Phi_{2}\vert^2 
   +w_{3,0}\vert\Phi_{1}^* \cdot \Phi_{2}\vert^2 
\nonumber
\end{eqnarray}
It is relevant for transitions where spin-density waves
play an important role, as, for example, in
high-$T_c$ superconductors (cuprates), see e.g. Refs.~\cite{ZDS-02,DPV-06}.

\item[$\bullet$]
Physically interesting examples of multicritical behavior
arise from the competition of orderings with symmetries O($n_1$) and
O$(n_2)$, at the multicritical point where the transition lines meet,
as shown Fig.~\ref{mcps}.  The corresponding LGW $\Phi^4$ theory must have
the symmetry O($n_1$)$\oplus$O($n_2$) and two O($n_1$) and O($n_2$)
vector fields $\phi_1$ and $\phi_2$,
\begin{eqnarray}
{\cal L}=(\partial_\mu \vec\phi_1)^2 
+(\partial_\mu \vec\phi_2)^2+ r_1 \vec\phi_1^{\,2}+ r_2 \vec\phi_2^{\,2}
+ u_1 (\vec\phi_1^{\,2})^2 + u_2 (\vec\phi_2^{\,2})^2 + 
w \vec\phi_1^{\,2} \vec\phi_2^{\,2} 
\label{mcp}
\end{eqnarray}
The multicritical behavior is observed by tuning two relevant scaling fields,
related to $r_1$ and $r_2$ \cite{KNF-76}.  Multicritical behaviors of this
type occur in several physical contexts: in anisotropic
antiferromagnets, high-$T_c$ superconductors, multicomponent polymer
solutions, etc..., see Refs.~\cite{KNF-76,CPV-03,HPV-05,PV-07}.

\end{itemize}

\subsection{Perturbative FT approach and RG flow}
\label{pertapproach}

The RG flow, and other interesting quantities as the critical exponents, 
can be determined by using perturbative approaches, extending the methods 
employed in the case of the O($N$) models, see
Sec.~\ref{perft}.
In the massive (disordered-phase) MZM scheme, one expands in powers
of the MZM quartic couplings $g_{ijkl}$, defined by
\begin{eqnarray}
\Gamma^{(2)}_{ij}(p) = \delta_{ij} Z_\phi^{-1} \left[ m^2+p^2+O(p^4)\right],
\qquad
\Gamma^{(4)}_{ijkl}(0) = m\,Z_\phi^{-2} \,g_{ijkl}
\label{mzmcom}
\end{eqnarray}
The massless (critical) $\overline{\rm MS}$ scheme is based on a minimal
subtraction procedure within the dimensional regularization, and can
give rise to an $\epsilon\equiv 4-d$ expansion, and also 3D expansions
in the renormalized $\overline{\rm MS}$ couplings $g_{ijkl}$ by
setting $\epsilon=1$ after renormalization.

High-order computations have been performed for several LGW $\Phi^4$ theories,
see e.g.
Refs.~\cite{KS-95,CPV-00,PS-00,PV-00,PRV-01,CPV-03,BPV-03,CP-04,CP-04-2,PV-05,BPV-05,DPV-06},
to five or six loops, which requires the calculation of approximately $1000$
Feynman diagrams. Again, the resummation of the series is essential. It can be
done by exploiting Borel summability and calculation of the large-order
behavior~\cite{CPV-00}.  This is achieved by extending the techniques employed
for O($N$) models.  The comparison of the results obtained from the analyses
of the MZM and $\overline{\rm MS}$ expansions provides nontrivial checks.

\begin{figure}
\centerline{\psfig{width=6truecm,angle=0,file=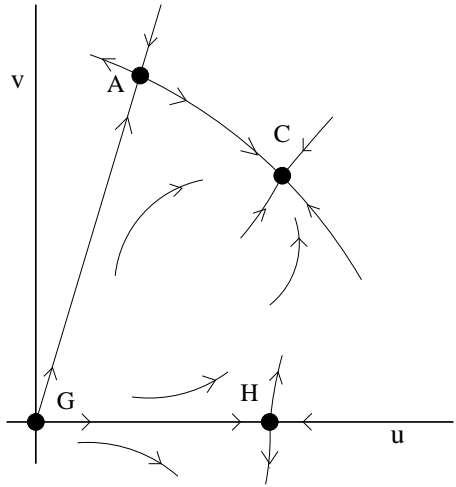}}
\caption{Example of RG flow in the coupling space of a two-parameter
$\Phi^4$ field theory. In particular, the figure 
sketches the RG flow in the renormalized coupling
$u$-$v$ plane of the
O($M$)$\otimes$O($N$) $\Phi^4$ theory in the large-$N$ limit, which
has four FPs: the Gaussian (G), O($M\times N$) (H), 
chiral (C) and antichiral (A) FPs; the stable FP is the chiral one.}
\label{rgflowomn}
\end{figure}

The RG flow is determined by the FPs, which are common zeroes $g^*_{ijkl}$ of
the $\beta$-functions, 
\begin{equation}
\beta_{ijkl}(g_{abcd})\equiv m {\partial g_{ijkl}\over \partial m}
\quad ({\rm MZM}), \qquad
\beta_{ijkl}(g_{abcd})\equiv \mu {\partial g_{ijkl}\over \partial \mu}
\quad (\overline{\rm MS}),
\label{betaf}
\end{equation}
in the MZM and $\overline{\rm MS}$ schemes respectively.  A FP is
stable if all eigenvalues of its stability matrix $S_{ij}=\partial
\beta_i/\partial g_j|_{g=g^*}$ have positive real part.
Fig.~\ref{rgflowomn} shows an example of RG flow in the
quartic-coupling plane of a two-quartic-parameter $\Phi^4$ field theory.

The existence of a stable FP implies that (i) physical systems with
the given global properties can undergo a continuous transition, (ii)
the asymptotic behavior in continuous transitions is controlled by the
stable FP (apart from cases requiring further tunings).  The absence
of a stable FP predicts first-order transitions between the disordered
and ordered phases of all systems.  Note that, even in the presence of
a stable FP, first-order transitions can be observed in systems
that are outside the attraction domain of the stable FP.

\subsection{The $\eta$ conjecture}
\label{etaconj}

Multi-parameter $\Phi^4$ theories have usually several FPs.  An
interesting question is whether a physical quantity exists such that
the comparison of its values at the FPs identifies the most stable FP.
In 2D unitary QFT, such a quantity is the central charge $c$, which is
related to the correlation function of the stress tensor
\begin{equation}
\langle T(z_1) T(z_2) \rangle = {c\over 2 z_{12}^4},
\label{cechT}
\end{equation}
and to the behavior of the free-energy of a slab, which is $f(L)|_{L\times
  \infty} = f_\infty L - c {\pi/(6L)}$ at $T=T_c$.  The
$c$-theorem~\cite{Zam-86} implies that the stable FP in a 2D unitary QFT is
the one with the least value of $c$.  But, despite several attempts and some
progress, see, for example, Refs.~\cite{cc2}, no conclusive results on the
extension of this theorem to higher dimensions has been obtained yet.

Within general $\Phi^4$ theories, the following conjecture has been put
forward~\cite{VZ-06}: {\em In general unitary $\Phi^4$ theories the infrared
  stable FP is the one that corresponds to the fastest decay of correlations.}
This corresponds to the FP with the largest value of the critical exponent
$\eta$ which characterizes the power-law decay of the two-point correlation
function $W^{(2)}(x)$ at criticality,
\begin{equation}
W^{(2)}(x) \propto {1\over x^{d-2+\eta}}.
\label{etadef}
\end{equation}
The exponent $\eta$ is related to the RG dimension of the field, $d_\Phi=(d-2+\eta)/2$. 
Since in $d<4$ general $\Phi^4$ theories may
have more than one stable FP with separate attraction domains, the $\eta$
conjecture should be then refined by comparing FP that are connected by RG
trajectories starting from the Gaussian FP: among them the stable FP is the
one with the largest value of $\eta$.

The $\eta$ conjecture holds in the case of the O($N$)-symmetric $\Phi^4$
theory.  For $d<4$, the Gaussian FP, for which $\eta=0$, is unstable against
the non-trivial Wilson--Fisher FP for which $\eta> 0$ (the positivity of
$\eta$ in unitary theories follows rigorously from the spectral representation
of the two-point function \cite{ZJ-book}).  It has been proven within the
$\varepsilon\equiv 4-d$ expansion, i.e. close to $d=4$.  It remains a
conjecture at fixed dimension $d<4$, but its validity has been confirmed by
several analytic and numerical results in lower dimensional cases.  
Several checks are reported in Ref.~\cite{VZ-06}.  No
counterexample has been found in all $\Phi^4$ theories studied so far.
This conjecture has also
been extended to $\Phi^4$ theories describing multicritical behaviors, such as
the O($n_1$)$\oplus$O($n_2$) $\Phi^4$ theory (\ref{mcp}).  In this situation, the
exponent $\eta$ is replaced by a matrix and the conjecture applies to the
trace of the matrix, i.e.  ${\rm Tr} \,\eta$.

\section{The O($M$)$\otimes$O($N$) $\Phi^4$ field theory}
\label{omnphi4th}

\begin{figure}
\centerline{\psfig{width=7truecm,angle=0,file=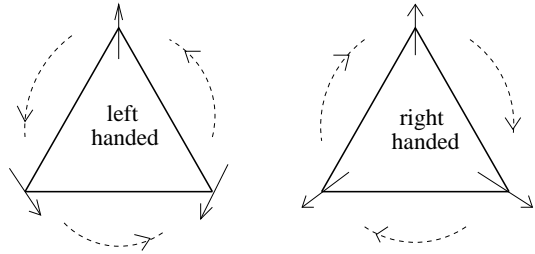}}
\caption{
The chiral 120$^o$ structure of the ground state of the
antiferromagnetic $XY$ model in a triangular lattice.
}
\label{chiralfig}
\end{figure}

In this section we discuss the RG flow of the O($M$)$\otimes$O($N$)
$\Phi^4$ theory (\ref{omnmo}). 
For $M=2$ and $v_0>0$, the symmetry breaking is
O(2)$\otimes$O($N$)$\rightarrow$O(2)$\otimes$O($N-2$), which may be roughly
simplified to O($N$)$\rightarrow$O($N-2$) (although the O(2) groups in the
symmetry-breaking pattern are not the same).  These cases describe transitions
in frustrated spin systems with noncollinear order~\cite{Kawamura-98}, where
frustration may arise either because of the special geometry of the lattice,
or from the competition of different kinds of interactions.  Typical examples
of systems of the first type are stacked triangular antiferromagnets (STA's),
for example CsMnBr$_3$, CsVBr$_3$, where magnetic ions are located at each
site of a three-dimensional stacked triangular lattice.  In STA's frustration
gives rise to a further degeneracy of the ground state, which presents a
chiral structure.  This can be easily seen in the two-component $XY$
antiferromagnetic model $H = \sum_{\langle xy \rangle} \vec{s}_x \cdot
\vec{s}_y$ defined on a triangular lattice.  The minimal energy configurations
have the chiral 120$^o$ structure shown in Fig.~\ref{chiralfig}.  Another
interesting  case is given by $M=2, N=3$, $v_0<0$,
which is relevant for the superfluid
transition of $^3$He, see e.g.  Ref.~\cite{DPV-04}.

The nature of the transition in frustrated systems with noncollinear order has
been longly debated.  Experiments on several physical systems show continuous
transitions, with exponents $\nu=0.57(3),0.54(3)$ for $N=2$ and $\nu=0.62(5)$
for $N=3$ (see e.g. Refs.~\cite{Kawamura-98,PV-r,DMT-04,CPPV-04} for
discussions of experimental results).  The existence of these new chiral
universality classes requires a stable FP in the 3-d RG flow of the
corresponding O(2)$\otimes$O($N$) $\Phi^4$ theory.  This issue was
investigated by using the $\epsilon\equiv 4-d$ expansion.  Calculations within
the $\epsilon=4-d$ expansion do not find any stable FP close to $d=4$, see
e.g.  Refs.~\cite{Kawamura-98,PV-r,DMT-04} and references therein.  The
extension of this result to the physical dimensions $d=3$ would predict
first-order transitions for any system, in apparent contradiction with
experiments. But the validity of such an extension to $d=3$ is not guaranteed,
because new FP's may appear going from $d\lesssim 4$ to $d=3$.
Fixed-dimension $d=3$ computations are required to check this possibility.

\begin{figure}[!bt]
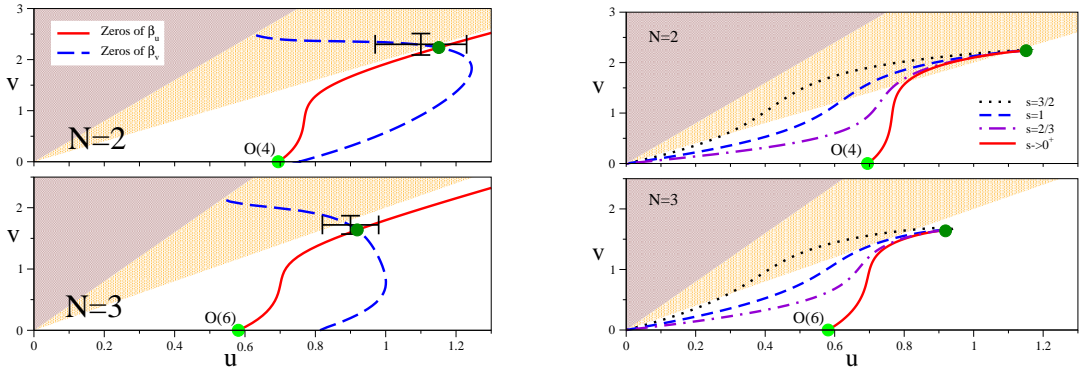

\begin{tabular}{cc}
\psfig{width=6.5truecm,angle=0,file=zeroes.eps} & $\qquad$ 
\psfig{width=6.5truecm,angle=0,file=flowOMN-MS.eps} \\
\end{tabular}
\vspace{2mm}
\caption{
  Some results for the RG flow of 3D O(2)$\otimes$O($N$) models obtained by
  the analysis of the five-loop $\overline{\rm MS}$ series~\cite{CPPV-04}.
  The left figure shows the zeroes of the $\overline{\rm MS}$ $\beta$
  functions. The right figure shows RG trajectories in the $u,v$ plane from
  the unstable Gaussian and O($2N$) FPs to the stable chiral FP for various
  values of the ratio $s\equiv v_0/u_0$ of the bare quartic parameters.}
\label{zerorgflow}
\end{figure}

The issue has been solved by high-order computations within 3D FT
schemes~\cite{PRV-01,CPS-02,CPPV-04}: 6-loop and 5-loop in the MZM and
$\overline{\rm MS}$ schemes respectively.  Their analyses provide a robust
evidence of the presence of a stable FP, supporting the existence of new 3D
chiral universality classes, which explain experiments. The left
Fig.~\ref{zerorgflow} shows the zeroes of the $\overline{\rm MS}$
$\beta$-functions.  The right Fig.~\ref{zerorgflow} shows the RG trajectories
in the $u,v$ plane from the unstable Gaussian to the stable chiral FP (which
can be obtained by solving the RG equation $- \lambda {dg_i/d\lambda} =
\beta_i(g_j)$, with $\lambda\in [0,\infty)$ and $g_j(0)= 0$, ${d g_i/
  d\lambda}|_{\lambda=0} = u_i$).  Critical exponents $\nu=0.57(3)$ for $N=2$
and $\nu=0.55(3)$ for $N=3$ are in substantial agreement with experiments.
The existence of these chiral universality class has been confirmed by MC
simulations of lattice models~\cite{CPPV-04,PS-03}.

The RG flow of the O($M$)$\otimes$O($N$) $\Phi^4$ theory can be also studied
analytically in the large-$N$ limit keeping fixed $M$, for any $2<d<4$.  In
the large-$N$ limit one finds four FPs as shown in Fig.~\ref{rgflowomn}, the
stable FP is the chiral one denoted by the letter $C$.  The values of the
critical exponent $\eta$ at the FPs provide a further confirm of the $\eta$
conjecture~\cite{VZ-06}, indeed setting $\eta = {\eta_1 e_d/N} + O(1/N^2)$
where $e_d$ is a constant depending on $d$, one finds~\cite{PRV-01-2}
$\eta_1=0$, $\eta_1=1/M$, $\eta_1=(M+1)/2$, and $\eta_1=(M-1)(M+2)/(2M)$
respectively for the Gaussian, O($M\times N$), chiral ad antichiral FPs.  All
numerical results at fixed $M,N$ satisfy the $\eta$ conjecture.

\section{Ferromagnetic transitions in disordered spin systems}
\label{rdis}

The FT approach allows us to also successfully describe ferromagnetic
transitions in disordered spin systems, which are of considerable theoretical
and experimental interest.  Such transitions are observed in spin systems
with impurities, such as mixing of antiferromagnetic materials with non
magnetic ones, for example Fe$_u$Zn$_{1-u}$F$_2$, Mn$_u$Zn$_{1-u}$F$_2$
(uniaxial), Fe$_{x}$Er$_z$, Fe$_{x}$Mn$_{y}$Zr$_z$ (isotropic), and also
$^4$He in porous materials.  
See e.g. Refs.~\cite{Belanger-00,PV-r,FHY-03,JBCBH-05} for 
experimental  and theoretical reviews.
These systems can be modeled by the lattice Hamiltonian
\begin{equation}
H_{\rho} 
= - J\,\sum_{\langle xy\rangle}  \rho_x \,\rho_y \;\vec{s}_x \cdot \vec{s}_y
\label{rdismodels}
\end{equation}
where the sum is over nearest-neighbor sites, $\vec{s}_x$ are $M$-component
spin variables, and $\rho_x=1,0$ with probability $p$ and $1-p$, respectively.
The disorder is quenched: it mimicks the physical situation in which the
relaxation time of the diffusion of impurities is much larger than other
typical scales. This implies that the free energy $F(\rho) \propto {\rm ln}
Z(\rho)$ must be averaged over the disorder. Accordingly, the expectation
value of an observable must be computed by
\begin{eqnarray}
\langle {\cal O} \rangle (\beta,\{\rho\}) = 
{\sum_{\{s\}} {\cal O} e^{-\beta {\cal H}(s;\rho)}
\over
\sum_{\{s\}} e^{-\beta {\cal H}(s;\rho)}
},
\qquad
\overline{\langle {\cal O} \rangle } = \int
[d\rho] P(\rho) \langle {\cal O} \rangle(\beta,\{\rho\})  
\nonumber
\end{eqnarray}
In the FT approach randomly-dilute spin models can be described by a $\Phi^4$
field theory for an $M$-component field $\Phi_i$ with and external random
field $\psi(x)$ coupled to the energy-density operator:
\begin{eqnarray}
{\cal L}_\psi =
{1\over 2} \sum_i (\partial_\mu \Phi_{i}(x))^2 + 
{1\over 2} (r+\psi(x)) \sum_i \Phi_{i}(x)^2  + 
{1\over 4!} g_0 (\sum_{i} \Phi_i(x)^2 )^2
\label{lpso}
\end{eqnarray}
where $\psi(x)$ is a spatially uncorrelated random field, with probability
$P(\psi)\sim {\rm exp} (-\psi^2/4 w)$.  Then, using the replica trick, ${\rm
  ln} Z = {\rm lim}_{n\rightarrow 0} (Z^n-1)/n$, one can formally integrate
over the disorder variables, arriving at a translation invariant $\Phi^4$
theory, which is the $MN$ model (\ref{mnmo}) with $u_0=-6w<0$.  The critical
behavior of the original system is determined by the RG flow of the $MN$ model
in the nonunitary limit $N\rightarrow 0$. Therefore, it can be determined by
analyzing the high-order MZM and $\overline{\rm MS}$ series for $N=0$, which
have been computed respectively to six loops~\cite{CPV-00,PV-00,PS-00} and to
five loops~\cite{KS-95,PV-05}.

\begin{figure}
\vspace{0.2cm}
\centerline{\epsfig{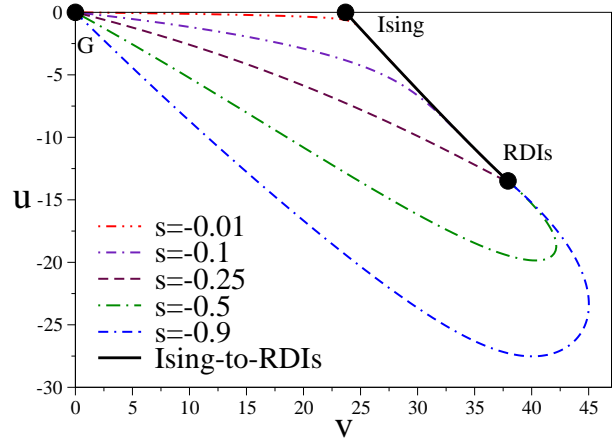}}
\caption{The RG flow of the $MN$ model 
  for $M=1$ and $N\rightarrow 0$ in the quartic-coupling plane.  It shows the
  trajectories from the unstable Gaussian and Ising FPs to the stable RDIs FP
  for various values of the ratio $s=u_0/v_0$ of the bare quartic parameters.
  Results obtained by the analysis of the six-loop MZM
  series~\protect\cite{CPPV-04-r}.}
\label{rgflowrdis}
\end{figure}

An interesting physical issue is whether the presence of impurities, and in
general of quenched disorder coupled to the energy density, can change the
critical behavior. General RG arguments~\cite{Harris-74,Aharony-76} show
that the asymptotic critical behavior remains unchanged if the specific-heat
exponent $\alpha$ of the pure spin system is negative, which is the case of
multicomponent O($M$)-symmetric spin systems, i.e. $M>1$.  On the other hand,
a different critical behavior is expected in the case of Ising-like systems
($M=1$), due to fact that $\alpha_{\rm Ising}=0.1096(5)$. This is confirmed by
the results of the analyses of the high-order FT perturbative series, which
show that the RG flow of the $MN$ model in the limit $N\rightarrow 0$ has a
stable FP in the region $u<0$ when $M=1$, see Fig.~\ref{rgflowrdis}, implying
the existence of a 3D randomly-dilute Ising (RDIs) universality class.

Experiments, see, e.g., Refs.~\cite{Belanger-00,PV-r}
and references therein, confirm this
scenario.  The asymptotic critical behavior remains unchanged
for multicomponent systems.
Table~\ref{RDIsres} reports results for Ising-like systems: from
experiments~\cite{Belanger-00}, the analysis of the six-loop FT
expansion in the MZM scheme~\cite{PV-00}, and recent Monte Carlo
simulations~\cite{HPPV-07}. The global agreement is very good.

\begin{table}
\begin{center}
\begin{tabular}{llll}
\hline
\multicolumn{1}{c}{RDIs exponents}& 
\multicolumn{1}{c}{$\nu$}& 
\multicolumn{1}{c}{$\beta$}\\ 
\hline  
EXPT \cite{Belanger-00}  & 0.69(1)   & 0.359(9) \\
PFT 6-$l$ MZM \cite{PV-00}  &  0.678(10) &  0.349(5)\\
MC \cite{HPPV-07}    & 0.683(2)  & 0.354(1) \\
\hline
\end{tabular}
\caption{
Results for the critical exponents of the 3D RDIs universality class.
}
\label{RDIsres}
\end{center}
\end{table}

It is worth mentioning that the RDIs universality class also describes
ferromagnetic transitions in the presence of frustration, when frustration is
not too large.  This is for exmaple found in the 3D $\pm J$ Ising
model~\cite{HPPV-07-2}, defined on a simple cubic lattice by the Hamiltonian
\begin{equation}
H_{\pm J} = - \sum_{\langle xy \rangle} J_{xy} \sigma_x \sigma_y, 
\quad  \sigma_x=\pm 1,
\label{pmj}
\end{equation}
where $J_{xy}=\pm J$ with probability $P(J_{xy}) = p \delta(J_{xy} - J) +
(1-p) \delta(J_{xy} + J)$.  This is a simplified model~\cite{EA-75} for
disordered and frustrated spin systems showing glass behavior in some region
of their phase diagram. Unlike model (\ref{rdismodels}), the $\pm J$ Ising
model is frustrated for any $p$. Neverthless, the paramagnetic-ferromagnetic
transition line extending for $1>p>p_N=0.76820(4)$ belongs to
the RDIs universality class, i.e., frustration turns out to be irrelevant at
this transition~\cite{HPPV-07-2}.  For $p<p_N$ the low-temperature phase is
glassy with vanishing magnetization, thus the critical behavior at the
transition belongs to a different Ising-glass universality class.

\section{The finite-temperature transition in hadronic matter}

The thermodynamics of Quantum Chromodynamics (QCD) is characterized by
a transition at $T\simeq 200$ Mev from a low-$T$ hadronic phase, in
which chiral symmetry is broken, to a high-$T$ phase with deconfined
quarks and gluons (quark-gluon plasma), in which chiral symmetry is
restored~\cite{reviews}.  Our understanding of the finite-$T$ phase
transition is essentially based on the relevant symmetry and
symmetry-breaking pattern. In the presence of $N_f$ light quarks the
relevant symmetry is the chiral symmetry ${\rm
SU}(N_f)_L\otimes{\rm SU}(N_f)_R $. At $T=0$ this symmetry is
spontaneously broken to SU($N_f$)$_V$ with a nonzero
quark condensate $\langle {\bar \psi} \psi \rangle$.  The finite-$T$
transition is related to the restoring of the chiral symmetry.  It is
therefore characterized by the simmetry breaking
\begin{equation}
{\rm SU}(N_f)_L\otimes {\rm SU}(N_f)_R   \rightarrow {\rm SU}(N_f)_V . 
\label{qcdsb}
\end{equation}
If the axial U(1)$_A$ symmetry is effectively restored at $T_c$, the
expected symmetry breaking becomes
\begin{eqnarray}
{\rm U}(N_f)_L\otimes {\rm U}(N_f)_R  \rightarrow {\rm U}(N_f)_V .
\label{qcdsbwa}
\end{eqnarray}
A suppression of the anomaly effects at $T_c$ is however unlikely in QCD.
Semiclassical calculations in the high-temperature phase~\cite{GPY-81} show
that instantons are exponentially suppressed for $T\gg T_c$, implying a
suppression of the anomaly effects in the high-temperature limit.  Some
lattice studies~\cite{anomalyMC} suggest a significant reduction of the
effective U(1)$_A$ symmetry breaking around $T_c$, but not a complete
suppression.  Since the anomaly, $\partial_\mu J_5^\mu \propto {1\over N_c}
Q$, gets suppressed in the large-$N_c$ limit, the symmetry-breaking pattern
(\ref{qcdsbwa}) may be relevant in the large-$N_c$ limit.

Other interesting QCD-like theories are ${\rm SU}(N_c)$ gauge theories
with $N_f$ Dirac fermions in the adjoint representation (aQCD). They
are asymptotically free only for $N_f < 11/4$, thus only the cases
$N_f=1,2$ are interesting.  Unlike QCD, aQCD is also invariant under
global ${\mathbb Z}_{N_c}$ transformations related to the center of
the gauge group SU($N_c$), as in pure ${\rm SU}(N_c)$ gauge
theories. There are two well-defined order parameters in the
light-quark regime, related to the confining and chiral modes,
i.e. the Polyakov loop and the quark condensate. Therefore, one
generally expects two transitions: a deconfinement transition at $T_d$
associated with the breaking of the ${\mathbb Z}_{N_c}$ symmetry, and
a chiral transition at $T_c$ in which chiral symmetry is restored. In
aQCD with $N_f$ massless flavors the chiral-symmetry group extends to
~\cite{SV-95} ${\rm SU}(2N_f)$. At $T=0$ this symmetry is expected to spontaneously
break to ${\rm SO}(2N_f)$, due to quark condensation.  Therefore the
symmetry breaking at the finite-$T$ chiral transition is 
\begin{equation}
{\rm SU}(2N_f) \rightarrow {\rm SO}(2N_f) 
\label{aqcdsb}
\end{equation}
with a  symmetric $2N_f\times 2N_f$ complex matrix as order parameter
related to the bilinear quark condensate.  If the axial U(1)$_A$
symmetry is restored at $T_c$, the symmetry-breaking pattern is
${\rm U}(2N_f) \rightarrow {\rm O}(2N_f)$.
MC simulations for $N_c=3$ and $N_f=2$ \cite{KL-99,EHS-05} show that
the deconfinement transition at $T_d$ is first order, while the
chiral transition appears continuous.
The ratio between the two critical temperatures turns
out to be quite large: $T_c/T_d\approx 8$.

\begin{table}
\begin{center}
\begin{tabular}{ccc}
\hline\hline
\multicolumn{1}{c}{}& 
\multicolumn{1}{c}{U(1)$_A$ anomaly}& 
\multicolumn{1}{c}{suppressed anomaly at $T_c$}\\
\hline
\multicolumn{1}{c}{QCD}& 
\multicolumn{1}{c}{${\rm SU}(N_f)_L\otimes{\rm SU}(N_f)_R\rightarrow{\rm SU}(N_f)_V$}& 
\multicolumn{1}{c}{${\rm U}(N_f)_L\otimes{\rm U}(N_f)_R\rightarrow{\rm U}(N_f)_V$}\\ 
\hline
$N_f=1$ & crossover or first order & O(2) or first order  \\
$N_f=2$ & O(4) or first order & 
U(2)$_L\otimes$U(2)$_R$/U(2)$_V$  or first order  \\
$N_f\ge 3$ & first order & first order \\
\hline\hline
\multicolumn{1}{c}{aQCD}& 
\multicolumn{1}{c}{${\rm SU}(2N_f)\rightarrow{\rm SO}(2N_f)$}& 
\multicolumn{1}{c}{${\rm U}(2N_f)\rightarrow{\rm O}(2N_f)$}\\ 
\hline 
$N_f=1$ & O(3) or first order & U(2)/O(2) or first order  \\
$N_f=2$ & SU(4)/SO(4) or first order  & first order \\
\hline\hline
\end{tabular}
\caption{Summary of the RG predictions.
We report the possible types of transition for each case,
indicating the universality class when the transition can also be
continuous.
}
\label{summary}
\end{center}
\end{table}

In order to study the nature of the finite-$T$ chiral transition in
QCD and aQCD, one can exploit universality and renormalization-group
(RG) arguments.~\cite{PW-84,BPV-03,BPV-05,BPV-05-l}

\begin{itemize}

\item[(i)]
Let us first assume that the phase transition at $T_c$ is continuous
for vanishing quark masses.  In this case the length scale of the
critical modes diverges approaching $T_c$, becoming eventually much
larger than $1/T_c$, which is the size of the euclidean ``temporal''
dimension at $T_c$.  Therefore, the asymptotic critical behavior must
be associated with a 3D universality class with the same symmetry
breaking. The order parameter must be an $N_f\times N_f$
complex-matrix field $\Phi_{ij}$, related to the bilinear quark
operators $\bar{\psi}_{Li}\psi_{Rj}$.

\item[(ii)] The existence of such a 3D universality class can be
investigated by considering the most general
LGW $\Phi^4$ theory compatible with the given symmetry breaking, which
describes the critical modes at $T_c$.  Neglecting the U(1)$_A$ anomaly,
it is given by
\begin{eqnarray}
{\cal L}_{{\rm U}(N)} = {\rm Tr} (\partial_\mu \Phi^\dagger) (\partial_\mu \Phi)
+r {\rm Tr} \Phi^\dagger \Phi 
+ {u_0\over 4} \left( {\rm Tr} \Phi^\dagger \Phi \right)^2
+ {v_0\over 4} {\rm Tr} \left( \Phi^\dagger \Phi \right)^2 .
\label{LUN}
\end{eqnarray}
If $\Phi_{ij}$ is a generic $N\times N$ complex matrix, the symmetry
is U$(N)_L\otimes$U$(N)_R$, which breaks to U$(N)_V$ if $v_0>0$, thus
providing the LGW theory relevant for QCD with $N_f=N$.  If
$\Phi_{ij}$ is also symmetric, the global symmetry is U$(N)$, which
breaks to O($N$) if $v_0>0$, which is the case relevant for aQCD with
$N_f=N/2$.  The reduction of the symmetry to
SU$(N_f)_L\otimes$SU$(N_f)_R$ for QCD [SU$(2N_f)$ for aQCD], due to
the axial anomaly, is achieved by adding determinant terms, such as
\begin{eqnarray}
{\cal L}_{{\rm SU}(N)} =  {\cal L}_{{\rm  U}(N)} +
w_0 \left( {\rm det} \Phi^\dagger + {\rm det} \Phi \right) .
\label{LSUN}
\end{eqnarray}
Nonvanishing quark masses can be accounted for by adding 
an external-field term ${\rm Tr} \left( H \Phi + {\rm h.c.}\right)$.

\item[(iii)] The critical behavior at a continuous transition is
determined by the FPs of the RG flow: the absence of a
stable FP generally implies first-order transitions.  Therefore, a
necessary condition of consistency with the initial hypothesis (i) of
a continuous transition is the existence of stable FP in the
corresponding LGW $\Phi^4$ theory.  If no stable FP exists, the finite-$T$
chiral transition of QCD (aQCD) is predicted to be first order.  If a
stable FP exists, the transition can be continuous, and its 
critical behavior is determined by the FP; but this does not exclude a
first-order transition if the system is outside the attraction domain
of the stable FP.

\end{itemize}

\begin{figure}
\centerline{\epsfig{width=7truecm,file=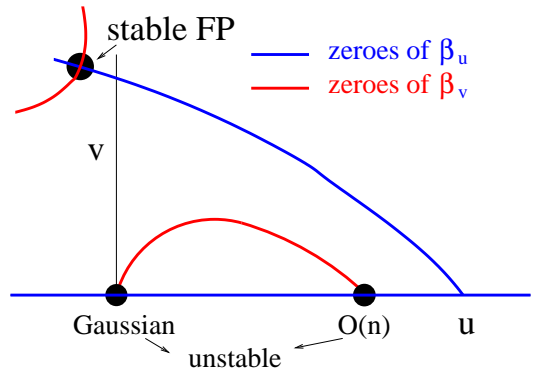}}
\caption{
Zeroes of the $\beta$-functions associated with the quartic
couplings of the Lagrangian (\protect\ref{LUN}) for $N=2$
(from Ref.~\cite{BPV-05-l}).
}
\label{zeroesu22}
\end{figure}

The above RG arguments show that the nature of the finite-$T$
transition in QCD and aQCD can be investigated by studying the RG flow
of the corresponding 3D LGW $\Phi^4$ field theories.  RG
studies based on high-order perturbative calculations in the MZM and
$\overline{\rm MS}$ are reported in
Refs.~\cite{BPV-03,BPV-05,BPV-05-l}.  Table~\ref{summary} presents a
summary of the predictions obtained by these RG analyses for the
finite-$T$ chiral transitions in QCD and aQCD with
$N_f$ massless quarks.

The case relevant to $N_f=1$ QCD, i.e. the Lagrangian (\ref{LUN}) with $N=1$,
reduces to the O(2) symmetric $\Phi^4$ theory, corresponding to the 3D $XY$
universality class.  The determinant term
related to the axial anomaly, cf. Eq.~(\ref{LSUN}), plays the role of an
external field, thus no continuous transition is expected, but a crossover.

In the case relevant for $N_f=2$ QCD with suppressed U(1)$_A$ anomaly, i.e.
the $\Phi^4$ theory (\ref{LUN}) with $N=2$, the analyses of both MZM and 3D
$\overline{\rm MS}$ schemes provide a robust evidence of a stable FP, see
Fig.~\ref{zeroesu22}.  A corresponding 3-D U(2)$\otimes$U(2)/U(2) universality
class exists, with critical exponents $\nu\approx 0.7$ and $\eta\approx 0.1$.
No stable FP is found close to $d=4$ by one-loop $\epsilon$-expansion
calculations~\cite{PW-84}, but, as already remarked in Sec.~\ref{omnphi4th},
the extension to $d=3$ of $\epsilon$-expansion results may fail.  A stable FP
is also found when the field $\Phi$ is symmetric, which is relevant for aQCD
with one flavor, thus showing the existence of a universality class
characterized by the simmetry breaking U(2)$\rightarrow$O(2).

\begin{figure}
\vspace{0.2cm}
\centerline{\epsfig{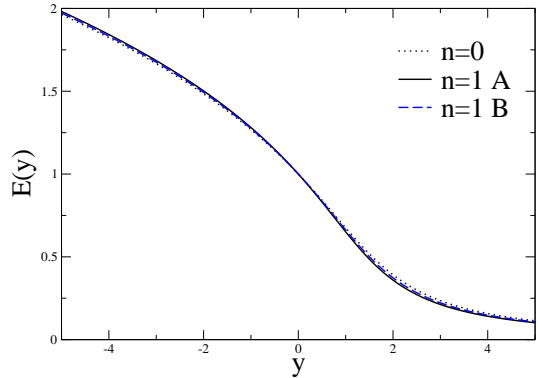}}
\caption{
The scaling function $E(y)$ for the O(4) universality class,
from Ref.~\cite{PPV-03}. The different lines represent different
approximations.
}
\label{eyo4}
\end{figure}

In two-flavor QCD, taking into account the U(1)$_A$ anomaly, the symmetry
breaking (\ref{qcdsb}) becomes equivalent to the one of the O(4) vector
universality class, i.e. O(4)$\rightarrow$O(3).  The symmetry breaking
(\ref{aqcdsb}) of $N_f=1$ aQCD is instead equivalent to the one of the O(3)
vector universality class.  This already suggests that, if the transition in the
chiral limit of $N_f=2$ QCD is continuous, it must show the same asymptotic
behavior of the 3D O(4) universality class (O(3) in the case of $N_f=1$ aQCD).
This implies that 
the critical equation of state of the 3D O(4) universality class, 
\begin{eqnarray}
\vec{M} \propto \vec{H} |H|^{(1-\delta)/\delta} E(y),
\qquad y\propto t |H|^{-1/(\beta+\delta)},
\label{o4rel}
\end{eqnarray}
provides a scaling relation between the quark condensate $\langle
\bar\psi \psi \rangle$, and the quark mass $m_f$, which correspond
respectively to the magnetization $M$ and the external field $H$.
The critical equation of state of the 3D O(4) universality class has
been accurately determined in the 3D O(4) vector model: the critical
exponents are~\cite{Has-01} $\delta=4.789(6)$, $\beta=0.3882(10)$, and
the universal scaling function $E(y)$ is shown in Fig.~\ref{eyo4}.

\begin{figure}
{\epsfig{width=15.2truecm,file=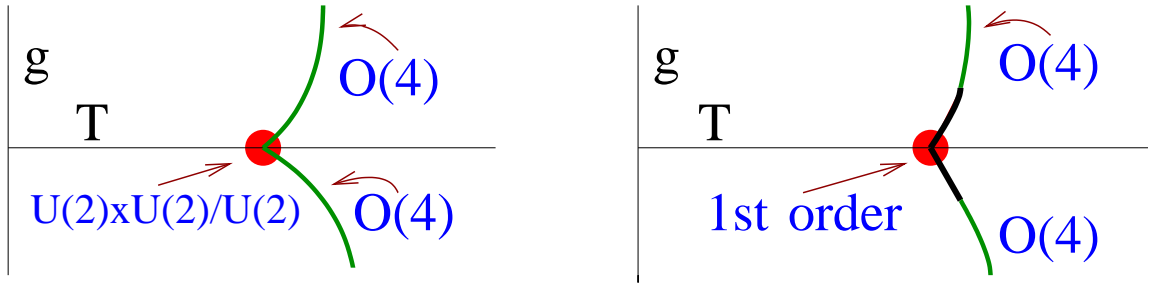}}
\caption{Possible phase diagrams in the $T$-$g$ plane 
for the LGW theory (\protect\ref{LSU2}) describing the transition of $N_f=2$ QCD,
in the case the transition at the multicritical point, i.e. for $g=0$,
is continuous (left) or first order (right).  Thick black lines
indicate first-order transitions. At their end points, thus for particular
values of $g$, the transition should be of mean-field type (apart from logarithms).
}
\label{pd}
\end{figure}

Actually, the LGW $\Phi^4$ theory corresponding to $N_f=2$ QCD 
is quite complicated~\cite{BPV-03,BPV-05-l}:
\begin{eqnarray}
&&{\cal L}_{{\rm SU}(2)} ={\rm Tr} (\partial_\mu \Phi^\dagger)
(\partial_\mu \Phi) +r {\rm Tr} \Phi^\dagger \Phi + {u_0\over 4}
\left( {\rm Tr} \Phi^\dagger \Phi \right)^2 + {v_0\over 4} {\rm Tr}
\left( \Phi^\dagger \Phi \right)^2+
\label{LSU2}\\
&&
+w_0 \left( {\rm det} \Phi^\dagger + {\rm det} \Phi \right)
+ {x_0\over 4} \left( {\rm Tr} \Phi^\dagger \Phi \right) 
         \left( {\rm det} \Phi^\dagger + {\rm det} \Phi \right) 
+ {y_0\over 4} \left[ ({\rm det} \Phi^\dagger)^2 + ({\rm det} \Phi)^2 \right] ,
\nonumber
\end{eqnarray}
where $w_0,x_0,y_0\sim g$ and $g$ parametrizes the effective breaking of the
U(1)$_A$ symmetry. If the anomaly is suppressed ($g=0$), then $w_0=x_0=y_0=0$.
${\cal L}_{{\rm SU}(2)}$ contains two quadratic (mass) terms, therefore it
describes several transition lines in the $T$-$g$ plane, which meet at a
multicritical point for $g=0$.  In the case of QCD the multicritical behavior
is controlled by the U(2)$_L$$\otimes$U(2)$_R$ symmetric theory.  Possible
phase diagrams in the $T$-$g$ plane are shown in Fig.~\ref{pd}.  When $g\neq
0$ the transition may be first order or continuous in the O(4) universality
class. Actually, we may also have a mean-field behavior (apart from
logarithms) for particular values of $g$, see Fig.~\ref{pd}.  If $|g|$ is
small (a partial suppression of anomaly effects around $T_c$ is suggested by
MC simulations~\cite{anomalyMC}) we may observe crossover effects controlled by
the U(2)$\otimes$U(2) multicritical point at $g=0$: if the transition is
continuous at $g=0$ then the free-energy should behave as ${\cal F}_{\rm sing}
\approx t^{3\nu} f(g t^{-\phi})$, where $t\propto T-T_c(g=0)$, $\nu\approx
0.7$, $\phi\approx 1.5$.  If the transition is first order at $g=0$, then it
is expected to remain first order for small $g$.  A similar scenario applies also
to $N_f=1$ aQCD.  

No stable FPs are found for $N>2$ in the LGW theory (\ref{LUN}).
Thus, neglecting the anomaly, transitions are always first order when
$N_f>2$ for QCD and $N_f>1$ for aQCD.  In most cases this result does
not change if we take into account the axial anomaly,
cf. Eq.~(\ref{LSUN}). The only exception is the case related to 
$N_f=2$ aQCD, where a stable FP is found~\cite{BPV-05}, corresponding
to a 3D SU(4)/SO(4) universality class, with critical exponents
$\nu\approx 1.1$ and $\eta\approx 0.2$.

In nature quarks are not massless, although some of them, the quarks $u$ and
$d$, are light. The physically interesting case is QCD with $N_f=2$ light
quarks and four heavier quarks (in particular the quark $s$ with $m_s\approx
100$ Mev). Therefore, it is important to consider the effects of the quark
masses in the above transition scenarios.  According to the above RG
arguments, if the transition is continuous in the chiral limit then an
analytic crossover is expected for nonzero values of the quark masses
$m_f$, because the quark masses act as external fields in the corresponding
LGW $\Phi^4$ theories.  On the other hand, a first-order transition is
generally robust against perturbations, and therefore it is expected to
persist for $m_f>0$, up to an Ising end point.  Actually, the presence of the
massive quark $s$ makes the above scenario more complicated, because the
nature of the transition may be sensitive to its mass $m_s$.  Since the
transition is expected to be first order in the chiral limit of $N_f=3$
degenerate quarks, we also expect that the first-order transition persists for
sufficiently small value of $m_s$.  On the other hand, if the transition is
continuous in the limit $m_s\rightarrow \infty$ corresponding to $N_f=2$
degenerate quarks, then there must be a tricritical point at $m_s=m_s^*$
(where the critical behavior is mean field apart logarithms) separating the
first-transition line from the O(4) critical line.

The nature of the transition in QCD can be investigated by lattice MC
simulations.  For $m_f>0$ around their physical values, MC simulations show
that the low-$T$ hadronic and high-$T$ quark-gluon plasma regimes are not
separated by a phase transition, but by an analytic crossover where the
thermodynamic quantities change rapidly in a relatively narrow temperature
interval, see, e.g., Refs.~\cite{Be-etal-05,Ch-etal-06,AFKS-06,FP-07,KFtalks}.
Neverthless, the nature of the transition in the chiral limit is still of
interest. Since the physical masses of the lightest quarks $u$ and $d$ are
very small, some scaling relations may still be valid at the physical values
of the quark masses, such as, for example, the O(4) relation (\ref{o4rel})
between the quark condensate and masses in the case the phase transition in
the chiral limit is continuous and belongs to the O(4) universality class.

The numerical investigation of the transition in the chiral limit is a hard
task because it must be studied in the infinite-volume limit
($V\rightarrow\infty$), in the continuum limit ($N_t\rightarrow\infty$ where
$N_t$ is the number of lattice spacings along the Euclidean time direction),
and massless limit ($m_f\rightarrow 0$).  A robust control of the scaling
corrections related to the continuum limit is essential.  One cannot even
exclude that in some lattice formulations the nature of the transition changes
when approaching the continuum limit. Moreover, in the case of a continuous
transition, the scaling corrections in the r.h.s. of Eq.~(\ref{freeen3}) may
hide the universal asymptotic behavior if MC simulations are not done
sufficiently close to the critical point (for example, in the case of the O(4)
universality class, the leading scaling-correction exponent is not large, i.e.
$\Delta\simeq 0.5$).

In the case of $N_f=2$ light quarks, the RG prediction for the chiral
transition is that it is first order or a continuous transition in the 3D O(4)
universality class. Many studies based on MC simulations of different lattice
formulations of QCD have been performed, see e.g.
Refs.~\cite{CP-PACS-01,MILC-00,KLP-01,KS-01,EHMS-01,DDP-05,KS-06}, but this
issue is still controversial. Some MC results favor a continuous transition.
However, the results have not been sufficient to settle its O(4) nature yet.
There are also results indicating a first-order transition.  Unlike $N_f=2$
QCD, the transition scenario appears settled for $N_f\ge 3$: MC simulations
\cite{KLP-01,IKSY-96,FP-03,Ch-etal-07,FP-07} show first-order transitions, in
agreement with the RG predictions.  Finally, in the case of $N_f=2$ aQCD the
available MC results \cite{KL-99,EHS-05} favor a continuous transitions. But
they are not yet sufficiently precise to check the critical behavior of the 3D
SU(4)/SO(4) universality class.

\end{document}